# Title

# Nanostructuring SiC by sequential plasma oxidation and reactive ion etching

## Authors

Joonas Isometsä*, Auguste Bieleviciute, Xiaolong Liu, Ville Vähänissi, Hele Savin

## Author contact information and affiliation

First author

Name: Joonas Isometsä (corresponding author)

Email: joonas.isometsa@aalto.fi

ORCID: https://orcid.org/0000-0003-2339-4586

Affiliation: Department of Electronics and Nanoengineering, Aalto University, Tietotie 3, 02150 Espoo, Finland

Second author

Name: Auguste Bieleviciute

Email: auguste.bieleviciute@aalto.fi

ORCID: https://orcid.org/0009-0006-6766-9157

Affiliation: Department of Electronics and Nanoengineering, Aalto University, Tietotie 3, 02150 Espoo, Finland & Fenno-Aurum Oy Ltd, Otakaari 5 A-GRID, 02150 Espoo, Finland

Third author

Name: Xiaolong Liu

Email: xiaolong.liu@aalto.fi

ORCID: https://orcid.org/0000-0002-1976-492X

Affiliation: Department of Electronics and Nanoengineering, Aalto University, Tietotie 3, 02150 Espoo, Finland

Fourth author

Name: Vähänissi Ville

Email: ville.vahanissi@aalto.fi

ORCID: https://orcid.org/0000-0002-2681-5609

Affiliation: Department of Electronics and Nanoengineering, Aalto University, Tietotie 3, 02150 Espoo, Finland

Fifth author (Professor)

Name: Hele Savin

Email: hele.savin@aalto.fi

ORCID: https://orcid.org/0000-0003-3946-7727

Affiliation: Department of Electronics and Nanoengineering, Aalto University, Tietotie 3, 02150 Espoo, Finland

# Data availability

The data that support the findings of this study are available from the corresponding author upon reasonable request.

# Acknowledgements

The authors thank Tandem Industry Academia, funded by the Finnish Research Impact Foundation, for funding the work. The work was also related to the Flagship on Photonics Research and Innovation "PREIN" funded by the Research Council of Finland, decision number 346529. The authors acknowledge the provision of facilities and technical support by the Micronova Nanofabrication Centre in Espoo, Finland, within the OtaNano research infrastructure at Aalto University. The authors would also like to express their gratitude to Fenno-Aurum for their valuable guidance and support throughout this work.

# Conflict of Interest

The authors declare no conflict of interest.

# Keywords

1. SiC
2. Nanostructuring

3. UV
4. ICP-RIE
5. Reflectance

# Abstract

Silicon carbide (SiC) is a highly promising material for the rapidly growing UV detection industry due to its visible-blindness, low dark current, and exceptional thermal and chemical stability. Despite these advantages, the performance of state-of-the-art SiC UV detectors remains limited due to high reflectance losses, even with the use of anti-reflection coatings. Here, we develop a reactive ion etching process for nanostructuring SiC to eliminate the reflectance losses. The process is based on consecutive oxidation and etching cycles. Consequently, a reflectance below 0.5% is achieved from deep UV (200 nm) to close to the SiC cut-off (~360 nm). The nanostructures are effective even at large incident angles as the reflectance remains practically unchanged up to 60 degrees. Furthermore, it is confirmed that the process consumes only ~1 µm of SiC and is compatible with $Al_2O_3$ masking, thereby facilitating straightforward integration into device fabrication. The developed cyclical etching process could also prove useful for SiC etching in general.

# 1. Introduction

UV detectors are a subject of wide interest across a large range of applications, including monitoring solar UV radiation, [1–3] flame detection, [4,5] and fields such as optical communication [6–8] and aerospace [9,10], just to name a few. Traditional photodiodes, such as those made of Si or Ge [11,12] are capable of detecting UV light; however, they are not visible-blind, i.e., they cannot separate the UV light from the visible light due to a too narrow bandgap or they require the use of complex filters, which results in considerably increased background noise or reduced signal level, respectively. Ideally, the material of choice for detecting UV would have a bandgap large enough that it would be transparent to the visible part of the solar spectrum, but small enough that UV light would be absorbed. This requirement is fulfilled by several semiconductors, including diamond, SiC, GaN, AlGaN, ZnS and ZnO, out of which particularly SiC has recently received an increased amount of attention due to its many benefits [6,13]. More specifically, the bandgap of SiC is 3.26 eV (4H-SiC), which enables UV detection across a broad wavelength range with a cut-off wavelength at 380 nm [14,15]. The large bandgap also naturally minimizes thermal generation, having a positive effect on UV sensitivity. Furthermore, SiC has exceptional temperature and chemical stability, along with excellent radiation hardness, making it suitable for a large variety of applications [14–16]. Additionally, high-quality SiC substrates are available in large quantities, and their price can be considered competitive in comparison to the other above-mentioned visible-blind materials, mostly due to their usage in mass-marketed power electronic devices.

Regardless of the attractive properties of SiC, its potential in UV detection has not been fully realized. Current SiC photodiodes suffer from considerable optical losses, namely surface reflectance. For planar SiC surfaces, the reflectance is around 25% between 200-350 nm [17]. A conventional method to mitigate the reflectance is to use anti-reflection coatings (ARC) that can reduce the reflectance to very low values. The downside is, however, that ARCs can be optimized only for a very narrow wavelength range [16,18,19]. As an example, the reflectance can be below 1% at a wavelength of 280 nm, while at 200 nm it reaches already 40% [18,20,21]. Additionally, ARCs suffer from performance loss at higher angles of incidence [22]. Another typical method to reduce reflectance is texturing the surface either using macroscopic (random pyramids or acidic texturing) or nanoscale structures, all of which have been realized for Si [23,24]. However, neither texturing method is used for SiC. Nanotexturing of SiC would be especially attractive as it would mimic an ideal antireflection treatment, where the refractive index changes gradually from air to that of SiC, eliminating reflectance-inducing interface. This effect could be achieved using a dense array of randomly distributed nanospikes [25,26] having dimensions in the range of UV photon wavelengths. In case of Si, several methods, including reactive-ion etching (RIE) [12,25], metal-assisted chemical etching (MACE) [27] and femtosecond laser patterning [28], are capable of forming such structures. However, such processes are not yet developed for SiC.

SiC is much more challenging to nanostructure than most single-element semiconductors due to its chemical stability and more complex surface chemistry [14,15]. Nevertheless, there are a few preliminary studies of SiC nanostructuring using RIE[17,29–31]. However, these studies focus primarily on the reduction of reflectance in the range of visible light [29–31]. Moreover, the approaches used are rather impractical and unfeasible for mass manufacturing as they rely on external masking layers such as thick and expensive epitaxial Si, slow electron beam lithography-based patterned mask, or difficult-to-remove and contamination-inducing Au nanoparticles [17,29,30]. Conversely, the complete opposite approach of SiC nanostructuring without any masking layer by relying on high ion bombardment, i.e., sputtering the surface, causes too much surface damage for the process to be useful at a larger scale [31]. A more practical approach, although a challenging one, would be to develop a process that would rely on the self-micromasking effect similar to the formation of nanostructured Si or Ge, for which the surface exposure to high-energy ion bombardment has been eliminated utilizing a versatile inductively coupled plasma reactive ion etching (ICP-RIE) system, resulting in negligible surface damage [12,25,32]. Furthermore, ICP-RIE systems would enable the usage of cyclical processes allowing both material deposition and etching in a consecutive manner during a single process. An example of such a process is the well-known Bosch etching of Si. Such a cyclical process could offer a clear advantage when considering the difficulties in etching SiC [14,33].

In this work, we develop a cyclical ICP-RIE process to form SiC nanostructures, aiming to eliminate reflectance losses from deep UV until the SiC cut-off wavelength around 380 nm. This process is based on two sequential steps: an in-situ oxygen plasma oxide masking step and an $SF_6/O_2$-based etching step, which are repeated for the predetermined number of cycles. The process performance is tested across the whole 100 mm diameter substrate area as well as on patterned wafers with etch mask openings of different sizes and shapes to mimic nanostructuring of actual devices. We investigate the morphology of the nanostructures as a function of the amount of cycles and characterize the optical performance via reflectance measurements at different angles of incidence. Furthermore, we use Raman spectroscopy to verify that the process does not change the crystallinity of the substrate. Finally, we discuss the potential of the developed process for etching SiC in general.

## 2. Background
### 2.1 Challenges in RIE of nanostructures on SiC
RIE of compound semiconductors is always more complex than single-element semiconductors as one needs to consider chemical reactions between all involved elements. In the case of SiC, RIE behaves similarly to the etching of Si and C atoms separately, given that the etching conditions are harsh enough to break strong Si-C bonds [14]. In these conditions, the Si removal rate is higher than that for C, and hence the RIE of SiC forms a thin carbon-rich surface layer consisting of mostly non-volatile $CF_x$ compounds, which limits the etching rate of the process [14,34]. An effective removal of these compounds typically requires physical bombardment (sputtering), which is also the most significant contributor to both the surface damage and the reduction of etching mask selectivity [14,35,36]. Therefore, it is important that the ion bombardment is minimized by adjusting it near the threshold of barely sufficient $CF_x$ removal. The most straightforward way to do this would be to use an ICP-RIE system, which has a separate plasma source for inductively coupled (ICP) and capacitively coupled (CCP) plasma, where the CCP accounts for most of the ion bombardment via its forward power, whereas the ICP determines the plasma density, which is a major factor for the etching rate [14]. ICP-RIE process development is typically quite complex due to the large number of individual process parameters, which often depend on each other, making the end result difficult to predict. A quite typical approach is just to experiment with the most impactful parameters such as the ratio of $SF_6$ to $O_2$, CCP and ICP plasma powers, process temperature, chamber pressure, and process duration.

Regarding the formation of nanostructures via RIE, a typical and well-proven approach is to utilize the micromasking effect. When etching the exposed area via plasma-excited etchant, e.g. $SF_6 + O_2$, some etching debris is created, which lands on the surface of the etched surface. This debris acts as a mask, resulting in random etching patterns. The formation of the patterns is accelerated by doing the process at cryogenic temperatures in the range of −100 °C to −150 °C. However, in the case of SiC, finding the balance between simultaneous micromasking and etching is much more challenging than for Ge or Si. Indeed, with the above approach, we were only able to create sparsely distributed micro/nanoscale structures that provided no noticeable optical benefits. These results motivated us to develop an alternative approach for SiC nanostructuring.

## 3. Results

### 3.1 The concept for cyclical oxidation and etching of SiC

Since the approach of simultaneous etching and masking did not result in the formation of nanostructures on SiC, we developed an alternative approach that relies on separated sequences. More specifically, this process relies on cycles that are divided into two steps – masking via oxidation and etching. In the first step (oxidation), SiC surface is oxidized via dense $O_2$ plasma (large ICP power, no CCP power) with the primary purpose of growing a masking $SiO_2$ layer on top of the SiC, followed by the additional effect of oxidizing carbon. This forms volatile $CO_x$ and hence reduces the amount of carbon residues on the surface [14]. In the second step (etching), the oxidized surface is exposed to etching gas consisting of $SF_6 + O_2$ with considerable CCP power, i.e., ion bombardment. This initiates the etching of the $SiO_2$ masking layer and consequently the SiC layer below it. The key to achieving nanostructures is to tailor the etching and the oxidation steps together such that the masking $SiO_2$ layer is locally fully removed during each cycle, i.e., some areas of bare SiC surface are exposed and etched, causing controlled surface roughening. This surface roughening is enhanced after each cycle, and it eventually leads to nanostructure formation. As a result, the nanospike height is mainly controlled by the number of cycles (duration), whereas the nanospike density, width and general shape are determined primarily by the process parameters in the etching step. The abovementioned process is summarized in Figure 1.

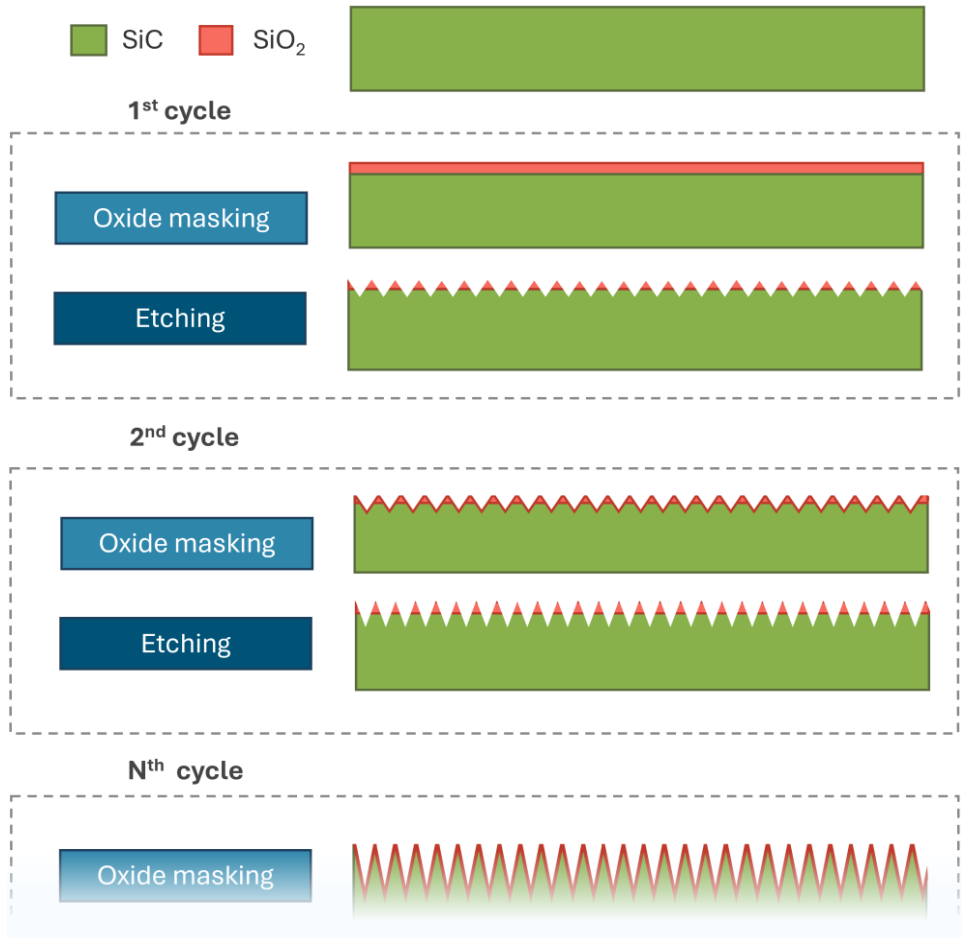

*Figure 1: A process diagram of the cyclical RIE nanostructuring process. Note that the whole process is carried out in ICP-RIE in one single recipe.*

### 3.2 Process optimization

As mentioned earlier, the nanostructure formation using the cyclical nanostructuring process heavily relies on the balance of oxidation and etching. The most straightforward method to control this balance is to adjust the duration of both oxidation and etching steps, assuming that the process parameters for both are set properly, i.e., the oxidation step oxidizes the surface and the etching step etches both $SiO_2$ and SiC. First, we set the substrate temperature to 60 °C in all steps in order to achieve the benefits of elevated temperatures (less etching residues and lower plasma power) without sacrifices in terms of process compatibility. Regarding the oxidation step, we wanted to make sure that the resulting oxide is thick enough for the etching and that oxidation does not induce any surface damage. Consequently, we set the ICP plasma power to a rather high value of 3 kW, combined with increasing the chamber pressure to 40 mTorr (tool limit). This resulted in a very dense plasma, while the DC bias remained near zero volts, indicating minimal ion bombardment that is optimal for a low-damage process. The baseline for the oxidation step duration was set to 1 minute, which should be enough for sufficient oxidation. In the etching step, we switched from this highly dense but gentle oxygen plasma to a moderately ion-bombarding plasma that can etch the SiC surface. This was done by adjusting the chamber pressure to 10 mTorr and decreasing the ICP power to 500 W, while increasing the CCP power to 65 W. This CCP power was optimized to a minimum value that still causes enough ion bombardment to initiate the etching process, yet low enough not to induce unnecessary surface damage. With these parameters, we found the balance between the oxidation and the etching when the etching duration was set to 9 s. More detailed process parameters are listed in Table 1 in the experimental section.

### 3.3 Resulting SiC nanostructures

After finding the balance for nanostructure formation, the next step is to study the formed nanostructures in more detail and to see how they evolve as a function of the cycle count. To be more precise, we tested 10, 20, 40 and 60 cycles and investigated the corresponding nanostructure morphology using SEM. The results are shown in Figure 2.

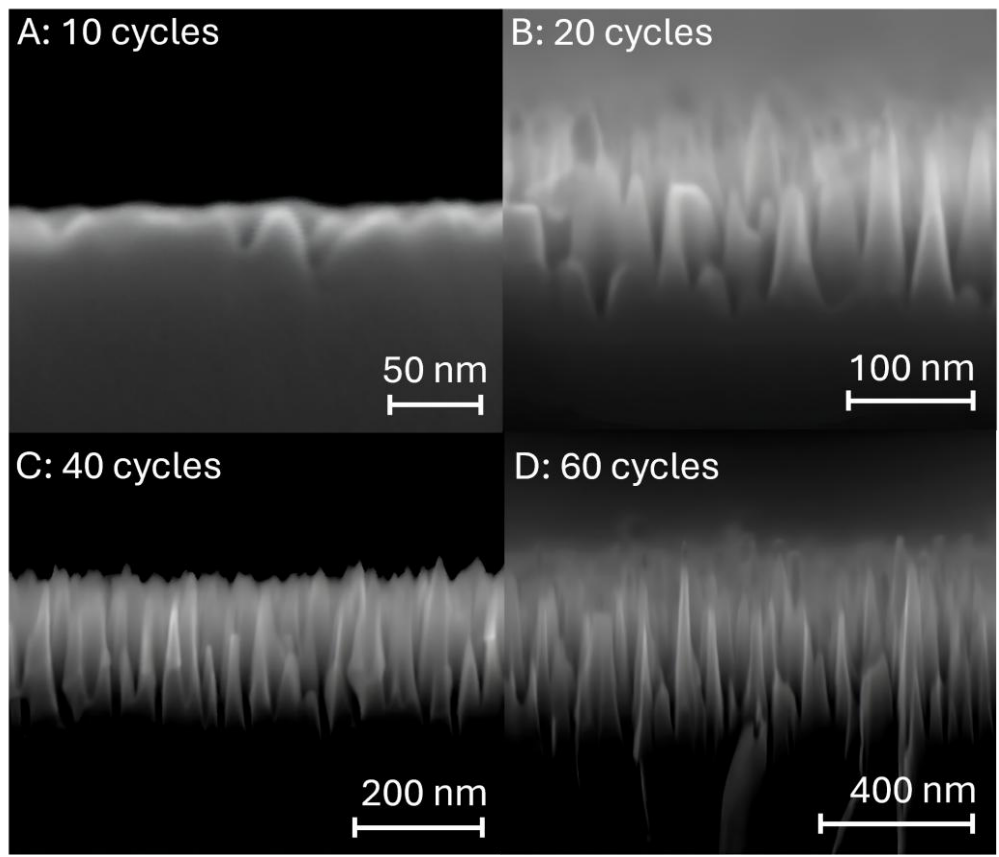

*Figure 2: Cross-sectional SEM images of SiC nanostructures resulting from (a) 10, (b) 20, (c) 40 or (d) 60 process cycles.*

The randomly distributed nanostructures are already visible after the first 10 cycles of the process (Figure 2a), although those nanospikes are still clearly underdeveloped, and the most visible spikes measure only 20-30 nm in both width (at the middle of the spike) and height. By increasing the cycle count to 20 (Figure 2b), there is a drastic change in surface morphology due to much more developed nanospikes that are in the size range of 80-100 nm in height and 20-30 nm in width (at the middle of the spike). However, there is still a considerable amount of underdeveloped nanospikes, indicating that the cycle count should be increased. Indeed, a similar trend continues since with 40 cycles (Figure 2c), the nanospikes are mostly fully developed, i.e., they all have a similar height of 200-220 nm, while the width remained similar at 20-30 nm. When the number of cycles was further increased to 60 (Figure 2d), the height of the nanospikes increased up to 400-500 nm, whereas their width remained the same. The nanospike height variation seems slightly higher in the case of 60 cycles compared to 40 cycles, which could be due to some nanospikes breaking during processing. Different aspect ratios achieved indicate that the process could be utilized in multiple applications, including those outside of optics. Furthermore, based on the SEM images, the spike sidewalls appear smooth, and no obvious signs of crystal damage/porosity could be observed.

## 3.4 Reflectance

Next, the optical properties of the same samples were characterized using a spectrophotometer with an integrating sphere, for a wavelength range of 250-350 nm. The corresponding results are shown in Figure 3a. As expected, the reference sample with a standard polished surface shows reflectance around 25% increasing towards lower wavelengths. After performing 10 cycles of the nanostructuring process, the reflectance is only slightly reduced, which is not too surprising considering that only minor changes were observed in surface morphology (Figure 2). By increasing the cycle count to 20 or more, the reflectance reduction becomes drastically more effective, and the reflectance drops to below 1% across the whole measured wavelength range. When inspecting the reflectance in more detail, the samples with either 40 or 60 cycles showed comparable and a bit lower reflectance than 20 cycles at below 0.5% reflectance, indicating saturation above 40 cycles. In summary, 20, 40, or 60 cycles all result in practically reflection-free surfaces independent of the wavelength, i.e., the nanostructures provide efficient broadband reflectance reduction.

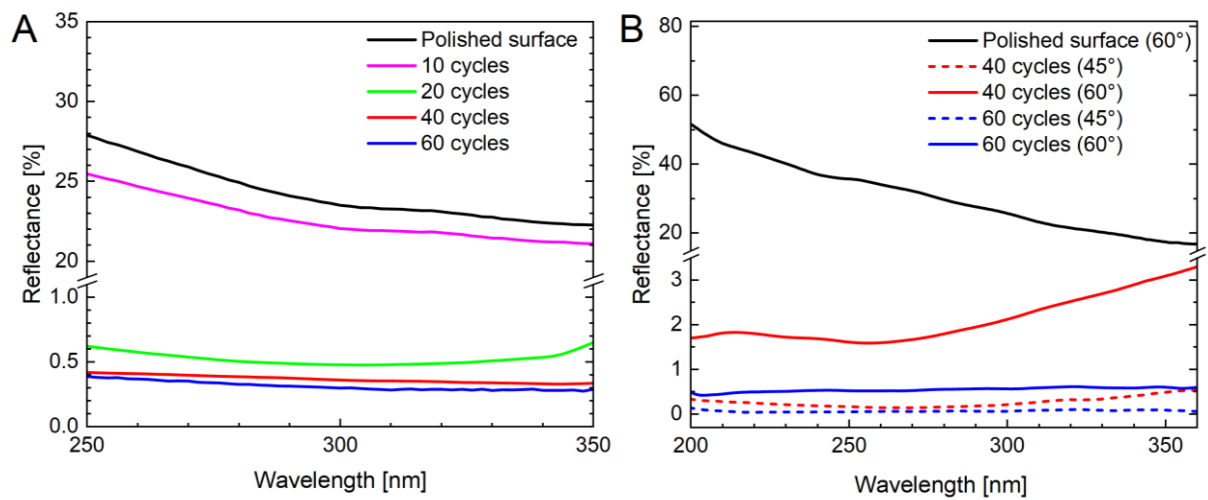

Figure 3: a) Reflectance of nanostructured (10, 20, 40 and 60 cycles) and polished SiC wafers in the wavelength range from 250 to 350 nm measured from perpendicular incident angle. b) Reflectance of nanostructured SiC wafers (40 and 60 cycles) at both 45° and 60° light incidence angles in the wavelength range from 200 to 350 nm. Polished SiC wafer with 60° light incidence angle shown as a reference. Note that there is a large axis break in the reflectance axis.

In many applications, the UV photons to be detected may arrive at the photodiode surface from various angles, not only from a perpendicular direction. Therefore, it is important to be able to reduce the reflectance independently of the incident angle. Consequently, we have measured the reflectance of selected samples (40 and 60 cycles) from two different incident angles (45° and 60°) and the results are shown in Figure 3b. As expected, the planar reference shows a significant increase in reflectance at higher incident angles (up to 50%). Similar behavior is expected for ARC samples [22]. In case of nanostructured samples, the reflectance remains very small at large incident angles in the case of both 40 and 60 cycle samples. More specifically, the sample with 40 cycles starts to show small deterioration at 60° (~3% reflectance), whereas the reflectance in the sample with 60 cycles remains practically unchanged (below 0.5%). This behavior can be explained by the gradually changing refractive index caused by the nanostructures, which reduces the angle dependency significantly [22].

Figure 4 compares the reflectance of the 60 cycle sample to the best results reported in the literature. These include a single layer ($SiO_2$) [18] and double layer ($Al_2O_3$ +$SiO_2$) [18] ARC samples as well as a nanostructured sample fabricated via electron beam lithography (EBL) [17]. As expected, ARC samples have a clear minimum reflectance in a narrow band, while the reflectance increases significantly outside this band. With a single-layer ARC, one cannot reach a similar level of reflectance than the nanostructured surfaces, whereas with a double-layer ARC, the minimum value is quite close to the values achieved with the nanostructuring process. The reflectance in the EBL reference sample is more uniform across the wavelengths, but remains much higher than in our sample. Moreover, it seems the reflectance would increase clearly towards lower wavelengths, although no data below 250 nm is available. It is worth highlighting that especially at the challenging deep-UV (200-250 nm) region, our nanostructure clearly outperforms all the earlier published results. In fact, our nanostructured surface shows no perceivable increase in reflectance even at 200 nm, and that, combined with the typical, gradually changing refractive index of a nanostructured surface, indicates that the exceptional antireflection properties of our nanostructured surface most likely extend well beyond 200 nm UV.

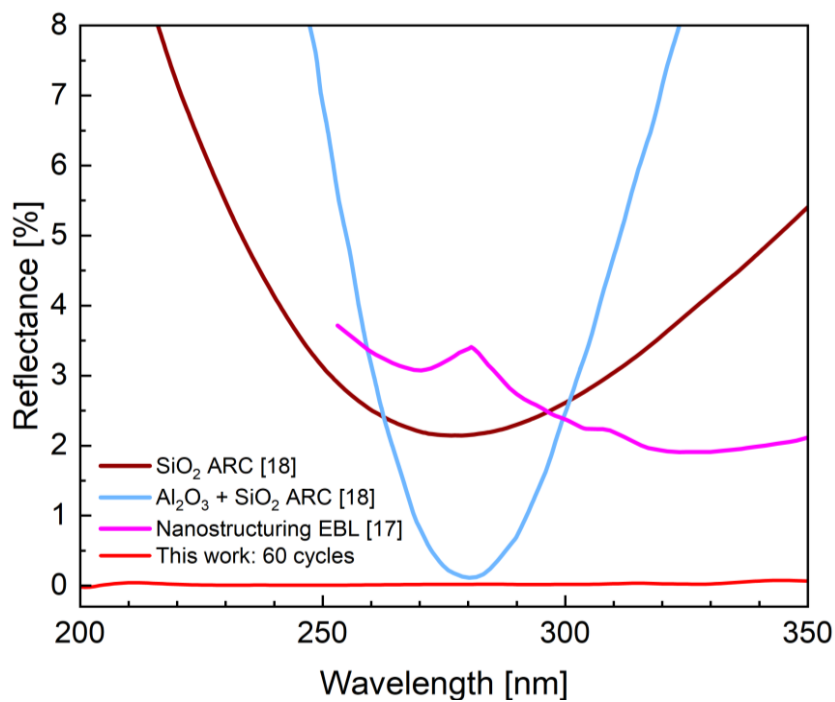

*Figure 4: Comparison of our 60 cycle sample to the best results reported in the literature. These include i) single layer ($SiO_2$) ARC [18], ii) double layer ($Al_2O_3$ +$SiO_2$) ARC [18] and iii) a nanostructured sample fabricated via electron beam lithography [17].*

## 3.5 Patternability and material consumption

Integration of our nanostructuring process to almost any practical device, including UV sensors, requires that it can be applied only to selected areas on the wafer. Furthermore, the etching result cannot depend on the size of that area, which is not trivial due to, e.g., the RIE loading effect [37].

Therefore, patternability and uniformity are of vital importance for device integration. In practice, patternability requires that the process is not so harsh that there would not be any reasonable masking material with high enough selectivity. Here, we tested an atomic layer deposited (ALD) $Al_2O_3$ masking layer with a thickness of 200 nm since $Al_2O_3$ is a widely available and well-known RIE masking material. After ALD, we patterned the $Al_2O_3$ masking layer with photolithography and wet etching. We varied the rectangular opening sizes from 0.0625 mm$^2$ to 100 mm$^2$. After mask layer patterning, 60 cycles of nanostructuring were applied. Visual inspection confirms that the $Al_2O_3$ mask works well, and nanostructures are formed only on the selected areas of the wafer. To determine the accurate reflectance values in the patterns, we measured the reflectance at selected single points inside and outside the patterned areas. At 250 nm, the values were 0.8% and 27% inside and outside the pattern, respectively. Furthermore, the mask opening size did not impact the reflectance values. To study the uniformity of the patterns, Figure 5 shows a reflectance map of the wafer. In the map, the patterned areas are clearly visible and homogenous across the wafer, independent of the opening size. It is worth noting that the reflectance map was measured at a wavelength of 855 nm due to the mapping tool limitations. As the reflectance with that wavelength is less than 1 %, the result confirms that our nanostructures also work at longer wavelengths.

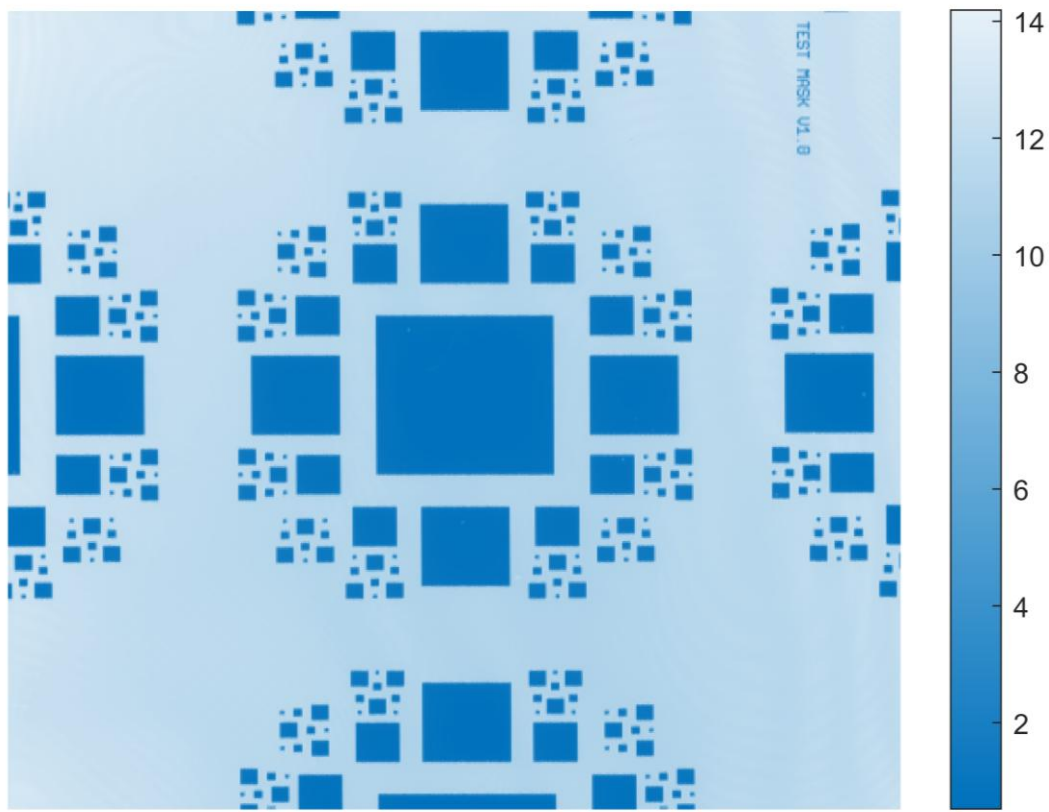

*Figure 5: a) Reflectance map from patterned SiC wafer with 60 cycles of nanostructuring process. The map combines both specular and diffuse reflectance at 855 nm wavelength.*

In devices, often only thin epitaxial SiC layers are used. These layers are typically at minimum a few micrometers thick, and hence, the amount of SiC consumed during the nanostructuring process becomes important [16,38–40]. Consequently, we studied the SiC consumption from the patterned sample, which was done by measuring the step height between masked and non-masked areas using both an SEM and a mechanical profilometer. Based on the characterization, the SiC consumption was 1 ± 0.1 µm, i.e., 16.5 ± 1.5 nm per cycle, which confirms that the process is compatible also with epitaxial SiC films. Additionally, SEM characterization confirmed that the $Al_2O_3$ mask eroded at a rate of 2 ± 0.5 nm per cycle.

3.6 Crystallinity

For optoelectronic devices, reduced reflectance is meaningless if the nanostructuring process simultaneously induces a high amount of surface damage or leaves behind large amount of etching residues that both lead to compromised electrical performance. To assess these possible issues, we performed Raman spectroscopy for the 40 cycle sample as the method is well known to be sensitive to lattice disorder, phase transitions, and chemical residues within a few micrometers of depth. The Raman was done on purpose directly after nanostructure formation, i.e., without any post-etching surface cleaning. The result is shown in Figure 6 together with a planar reference.

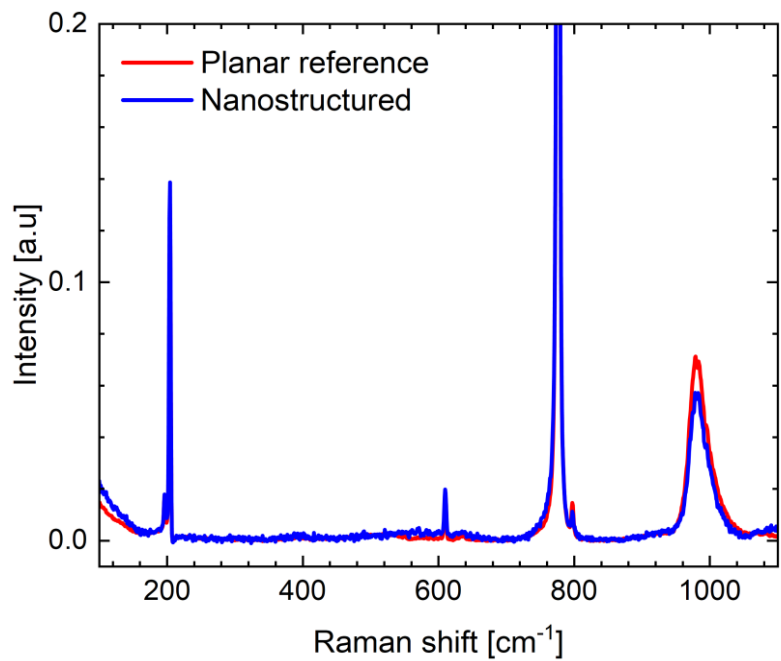

*Figure 6: A comparison of normalized Raman spectra from nanostructured and planar reference SiC wafers.*

The Raman spectrum from the nanostructured sample shows no significant peak broadening or peak shifts compared to the planar reference sample, indicating no notable changes in disorder or stress in the surface crystal structure [41]. Furthermore, both samples share the same peaks, which are all known SiC peaks, meaning that no notable Si-Si, Si-F, C-C, or C-F peaks are visible [42–46]. This suggests that there are no apparent changes in surface bonds or a significant amount of etching residues remaining. While the confocal Raman spectroscopy used here can detect major surface damage, it is limited in depth resolution, typically on the order of hundreds of nanometers to a few micrometers, and its volumetric probing nature makes it challenging to detect ultrathin damage layers (e.g., <10 nm) or regions with low defect density. Such subtle damage could still affect electrical properties, motivating further studies focused on nanoscale to atomic-scale structural defects and the use of more electrically sensitive characterization techniques. Additionally, while the Raman result does not fully exclude the presence of small amounts of etching residues, if there are some, standard RCA cleaning could be done to remove most of them. Hence, we confirmed that the nanostructures tolerate RCA cleaning well, i.e., the reflectance and morphology remained in practice the same before and after the cleaning. This was not self-evident, as in the past, it has been shown that RCA can affect the optical properties of nanostructured silicon.[47]

## 4. Discussion

A natural first application for testing our nanostructures in actual devices would be UV sensors. If we consider their sensitivity and how the reduced reflectance after the nanostructuring could affect it, the most relevant device parameter to focus on is the external quantum efficiency (EQE). EQE describes the number of output electrons from the sensor for every incident photon. In other words, it can be considered as the probability at which each incoming photon is detected. It is naturally determined by two factors: i) reflectance and ii) recombination. The former affects the amount of light entering the device, while the latter affects the amount of charge carriers reaching the outer circuit. Since state-of-the-art SiC PIN UV photodiodes rely on traditional ARCs, the EQE will decrease considerably outside the optimized narrow wavelength band. For instance, a SiC photodetector with $Al_2O_3$ + $SiO_2$ ARC has above 40% reflectance at wavelengths 200 – 210 nm [20], and hence has the achievable maximum EQE value less than 60%. In this work, by nanostructuring the surface, we were able to reduce the reflectance below 0.5% from a wavelength range of 200 – 360 nm, i.e., deep UV until near SiC cut-off. Hence, with the nanostructure, the achievable maximum EQE would stay very close to the ideal 100% value. Of course, as mentioned above, one needs to keep in mind that EQE does not depend only on the reflectance but also on bulk and surface recombination. Therefore, even though our preliminary findings indicate no damage at the surface after nanostructuring, the recombination parameters, including efficient surface passivation, must be confirmed before quantitative estimations for EQE can be made.

As mentioned earlier, plasma etching of SiC is known to be challenging, often resulting in undesired RIE induced surface damage, roughness and microtrench formation [14]. Moreover, the incomplete understanding of the underlying etching mechanisms and the difficulty of accurately predicting etch rates complicate process optimization [14]. Therefore, alternative etching approaches would likely be highly appreciated by the SiC community. In this paper we have focused on the formation of nanostructures and developed a cyclical RIE process for that. However, the concept of our cyclic RIE process could potentially be utilized in SiC plasma etching in general as well. The process developed here could have several similar benefits than those reported earlier for the so-called CORE process developed for silicon [48,49]. Firstly, the masking layer growth should be self-limiting due to the logarithmic growth rate of the oxide [48,49]. Secondly, the masking layer should be uniform, allowing the formation of even higher aspect ratio structures and improved consistency on various-sized patterns [48,49]. Thirdly, the oxide mask does not accumulate on samples or processing tools, thereby minimizing etching residues and reducing process-related inconsistencies [48,49]. Nevertheless, a further set of experiments is needed to confirm this and reveal more about the potential.

## 5. Conclusions

In this work, we have successfully fabricated nanostructures on SiC surfaces. This was achieved by developing a cyclical process based on sequential plasma oxidation and reactive ion etching. The nanostructured surface consists of a dense array of randomly distributed nanospikes with dimensions comparable to UV photon wavelengths. Furthermore, the dimensions of the fabricated nanospikes could be easily controlled by simply adjusting the number of process cycles. The nanostructured SiC substrate showed both uniform and exceptionally low (below 0.5%) reflectance throughout the measured wavelength range (200 – 360 nm). Furthermore, the reflectance remained minimal even with large incident angles up to 60°. Additionally, the process left neither significant signs of surface damage nor etching residues, and it proved to be compatible with standard lithography-based patterning when using $Al_2O_3$ as the masking material. Moreover, the process provided the same optical properties regardless of the mask opening size, demonstrating excellent versatility for further device integration. It also consumed only a small amount of bulk SiC (1 μm), confirming the compatibility with thin epitaxially grown SiC layers. Based on the achieved results, the nanostructuring process has a potential to considerably improve the performance of SiC UV photosensors.

## 6. Experimental section

### 6.1 Sample fabrication

A batch of 350 μm-thick, double-side polished 100 mm diameter n-type {0001} with 4 degree off-axis 4H-SiC wafers with a 0.015-0.028 ohm-cm base resistivity were used in these experiments. Before any processing, the wafers were cleaned with both SC-1 and SC-2. Surface nanostructuring was formed using inductively-coupled plasma reactive ion etching (ICP-RIE) (Oxford Instruments PlasmaPro 100 Estrelas) with helium backing sample cooling/heating with either a chiller or liquid nitrogen. Samples were heated to the process temperature in-situ before starting the run. The etching process contained two different steps, namely oxidation and etching, which were repeated sequentially by the specified number of cycles (10-60 in this study). Table 1 summarizes the key process parameters for each step.

Table 1 Process parameters for the cyclical ICP-RIE-based process for SiC nanostructuring. Each cycle consists of oxidation and etching steps, which are repeated for the specified number of cycles.

| Parameter | Oxidation | Etch |
| --- | --- | --- |
| Chamber pressure | 40 mTorr | 10 mTorr |
| $O_2$ gas flow | 50 sccm | 15 sccm |
| $SF_6$ gas flow | 0 sccm | 35 sccm |
| Temperature | 60 °C | 60 °C |
| Plasma power ICP | 3000 W | 500 W |
| Plasma power CCP | - | 65 W |
| CCP plasma frequency | 0 Hz | 400 kHz |
| Duration | 1 min | 9 s |

For patterning tests, some wafers were coated with 200 nm (2000 cycles) of thermal ALD $Al_2O_3$ deposited at 160 °C using Picosun SUNALE R-200 Advanced tool. The $Al_2O_3$ layer was patterned to contain various-sized square openings, ranging from 0.0625 mm² to 100 mm², mimicking actual optical devices. The patterning was done using standard photolithography and wet etching in BHF. This was then followed by performing 60 cycles of the nanostructuring process, resulting in treated surface in the exposed areas.

### 6.2 Sample characterization

The surface morphology was inspected by cross-sectional scanning electron microscopy using a FEI Helios Nanolab 600 tool with electron acceleration voltage and beam current set to 5 kV and 0.17-0.69 nA, respectively. The spectroscopic measurements of reflectance for a fixed angle of incidence were measured in-house with an Agilent Cary 5000 spectrophotometer (Figure 3) or Bentham PVE300 (250 nm wavelength single point measurement) with an integrating sphere, and the spectroscopic measurements at varying angles of incidence were conducted using an Agilent Cary 7000 spectrophotometer with a universal measurement accessory by an external service provider (Measurlabs, Helsinki). The reflectance map measurements were conducted with a Semilab PV2000A tool using a 0.1 mm raster resolution and an 855 nm laser. All reflectance measurements here include both specular and diffuse reflectance with a reasonable amount of averaging combined with curve smoothing to reduce visible measurement noise. The Raman spectra were measured with an excitation wavelength of 532 nm using a WiTec Alpha300 RA+ tool. The amount of SiC consumed by the process in the vertical direction was determined by profilometry (Dektak XT) and verified with SEM.